\documentclass[epj]{svjour}
\usepackage{graphics}
\usepackage[pagewise]{lineno}
\begin{document}
\def\prl{{\em Phys. Rev. Lett. }}
\def\epja{{\em Eur. Phys. J.  A }}
\def\epl{{\em Europhys. Letts.}}
\def\prc{{\em Phys. Rev.  C }}
\def\prd{{\em Phys. Rev.  D }}
\def\jap{{\em J. Appl. Phys. }}
\def\ajp{{\em Am. J. Phys. }}
\def\nima{{\em Nucl. Instr. and Meth. Phys. A }}
\def\npa{{\em Nucl. Phys. A }}
\def\npb{{\em Nucl. Phys.  B }}
\def\npbp{{\em Nucl. Phys.  B (Proc. Suppl.) }}
\def\njp{{\em New J.  Phys. }}
\def\epjc{{\em Eur. Phys. J.  C }}
\def\plb{{\em Phys. Letts. B }}
\def\phy{{\em Physics }}
\def\mpla{{\em Mod. Phys. Lett. A }}
\def\prep{{\em Phys. Rep. }}
\def\zpc{{\em Z. Phys. C }}
\def\zpa{{\em Z. Phys. A }}
\def\ppnp{{\em Prog. Part. Nucl. Phys. }}
\def\jpg{{\em J. Phys. G }}
\def\cpc{{\em Comput. Phys. Commun. }}
\def\app{{\em Acta Physica Pol. B }}
\def\aip{{\em AIP Conf. Proc. }}
\def\jhep{{\em J. High Energy Phys. }}
\def\psc{{\em Prog. Sci. Culture }}
\def\snc{{\em Suppl. Nuovo Cimento }}
\def\sjnp{{\em Sov. J. Nucl. Phys. }}
\def\ptp{{\em Prog. Theor. Phys. }}
\def\ptps{{\em Prog. Theor. Phys. Suppl. }}
\def\iansf{{\em Izv. Akad. Nauk: Ser. Fiz. }}
\def\jpcs{{\em J. Phys. Conf. Ser. }}
\def\aipcp{{\em AIP Conf. Proc. }}
\def\ahep{{\em Adv. High Energy Phys. }}
\def\ijmpe{{\em Int.  J. Mod. Phys. E }}

\def\jour#1#2#3#4{{#1} {\bf#2} (19#3) #4}
\def\jou2#1#2#3#4{{#1} {\bf#2} (20#3) #4}
\def\ct{\cite}
\def\sNN{\sqrt{s_{NN}}}
\def\sNNq{s_{NN}}
\def\spp{\sqrt{s_{pp}}}
\def\eNN{\varepsilon_{NN}}
\def\pbp{{\bar p}p}

\def\col{Collab.}
\def\bi{\bibitem}
\def\ea{{\sl et al.}}
\def\eg{{\sl e.g.}}
\def\vrs{{\sl vs.}}
\def\ie{{\sl i.e.}}
\def\va{{\sl via}}
%


\title{ Effective energy 
 budget in
 multiparticle production in 
 nuclear
collisions}

\author{\large Aditya Nath Mishra$^{1,}$\thanks {e-mail: Aditya.Nath.Mishra@cern.ch}, 
\large Raghunath Sahoo$^{1,}$\thanks {e-mail: Raghunath.Sahoo@cern.ch}, 
\large Edward K.G. Sarkisyan$^{2,3,}$\thanks {e-mail: sedward@mail.cern.ch}, 
\large and Alexander S. Sakharov$^{2,4,}$\thanks {e-mail: Alexandre.Sakharov@cern.ch }}
\institute{Discipline of Physics, School of Basic Sciences, Indian Institute of Technology Indore, Indore-452017, India \and Department of Physics, CERN, 1211 Geneva 23, Switzerland \and Department of Physics, The University of Texas at Arlington, Arlington, TX 76019, USA \and 
Department of Physics, Kyungpook National University, Daegu 702-701, 
Korea}

%
%
%
\date{Received: date / Revised version: date}
%
\abstract{
 The dependencies of charged particle 
pseudorapidity density and 
transverse energy  
pseudorapidity density at 
midrapidity on the collision energy and on the number of nucleon 
participants, or centrality, measured in nucleus-nucleus collisions are 
studied in the energy range spanning a few GeV to a few TeV per nucleon.
 The 
 approach
 in 
 which 
 the multiparticle production 
 is driven  
  by the dissipating effective energy of participants 
 is introduced.
 This approach
  is 
 based on the earlier proposed 
 consideration,
 combining the constituent 
quark picture together with Landau relativistic hydrodynamics shown to 
 interrelate the measurements
 from different types of 
collisions.
 Within this 
 picture,
 the 
 dependence 
 on 
 the number of participants in  heavy-ion collisions 
 are found to be well described in terms of the effective energy defined 
as a 
 centrality-dependent fraction of 
 the collision energy.
 For both variables 
 under study,
the effective energy approach  reveals
a 
similarity in the energy 
dependence  obtained for the most central collisions and 
centrality data 
 in the entire available energy range.
 Predictions are made for the investigated dependencies 
 for 
 the forthcoming  
 higher energy
 measurements in heavy-ion  
  collisions 
 at the LHC.
%
} 
\maketitle
%
{\bf 1.} Multiparticle production in high-energy particle and nuclear 
collisions 
attracts high interest, as, on the one hand, 
the observables measured 
first in high-energy collisions, namely multiplicity and transverse energy, 
are immediate characteristics of this process and bring 
important information on the underlying dynamics of strong interactions, 
while on the other hand,  
this process still eludes its 
complete understanding.  
 It is already more than half a century as the multiplicity of the produced
particles are considered to be derived by the collision energy
\ct{Heisenberg,Fermi,Landau}.  In this picture the energy pumped into 
 the collision zone in the very first stage of the collision defines the 
 volume of the 
 interaction lump 
 of 
participant patterns. Later on, the 
approach of ``wounded'' nucleons, or nucleon participants,  has been 
proposed to describe the 
multiplicity and particle distributions \ct{wound}. In this 
approach the multiplicity is expected to be proportional to the number of 
participants. However it was 
observed at RHIC and similarly at LHC energies, the 
concept of wounded nucleons   
does not describe the measurements where the data found to demonstrate 
an increase with the number of nucleon participants.         The problem 
has 
been addressed in the nuclear overlap model using Monte Carlo 
simulation in the constituent quark 
framework, and the scaling has been shown to be restored 
\ct{eremin,ind,indpart,nouicer}. In addition, it was observed that the 
multiplicity and 
midrapidity-density distributions are similar in 
 $e^+e^-$
  and in the 
most central (head-on) nuclear collisions \ct{phobos-sim} at the same 
center-of-mass 
 (c.m.) energy
 pointing to 
  the universality of multihadron production.   
However, the expectation to 
 observe this type of universality 
 in hadronic 
and nuclear collisions at similar
c.m. energy per nucleon has not been shown by the data where the 
measurements in hadron-hadron collisions 
 show 
significantly lower values
 compared to 
 those 
 in central heavy-ion collisions \ct{book,advrev}.    

To interpret these observations, the energy dissipation 
 approach
 of 
constituent quark participants has been proposed in \ct{edward} by two of 
the 
authors of this paper.
 In this 
 picture,
 the process of 
particle production is 
 driven by
 the 
 amount of 
 energy 
 deposited
 by 
interacting 
participants into the small Lorentz-contracted volume during the early 
stage of the collision. The whole process of a collision  
is then considered as the expansion and the subsequent break-up into 
particles from an initial state. This approach resembles the Landau 
phenomenological hydrodynamic approach  of multiparticle 
production in 
relativistic particle interactions \ct{Landau}, which was found to be in a 
good 
agreement with the multiplicity data in particle and nuclear collisions in 
the wide energy range. In 
the 
 picture
proposed in \ct{edward}, the Landau hydrodynamics  is combined
with the constituent quark 
 model
\ct{constitq}.
 This makes 
 the secondary particle  production 
 to be  
 basically driven by 
 the amount of the initial {\it effective} energy deposited by 
participants -- quarks or nucleons, into the Lorentz contracted overlap 
region. In 
$pp/\pbp$ collisions, a single constituent (or 
dressed) quark from each nucleon takes part in a collision and rest are 
considered as spectators. 
Thus, the effective energy for the production of secondary 
particles is the energy carried by a single quark pair \ie\ 1/3 of 
the entire 
nucleon energy. 
 In contrary,
 in the head-on heavy-ion collisions, 
 the participating nucleons 
 are considered colliding by 
 all three 
constituent quarks from each nucleon 
 which makes  
 the 
 whole 
 energy of the colliding nucleons (participants) 
 available for 
 secondary
 particle production. 
 Thus, one 
 expects that 
 bulk 
observables 
 measured 
 in the 
 head-on
 heavy-ion collisions at 
the c.m. energy per nucleon, $\sqrt{s_{NN}}$,
to be 
similar to those from  $pp/\pbp$ collisions but at a 
 three times larger 
 c.m. energy \ie\ $\sqrt{s_{pp}} \simeq 3\sqrt{s_{NN}}$.

 Combining the above discussed ingredients of the constituent quarks 
and  Landau hydrodynamics, one obtains 
 the relationship between charged particle rapidity 
 density per participant 
 pair, $\rho(\eta)=(2/N_{\rm{part}})dN_{\rm{ch}}/d\eta$ at midrapidity 
 ($\it{\eta} \approx $~0) in heavy-ion collisions and that 
 in $pp/\pbp$ collisions: 
 \begin{equation}
 \frac{\rho(0)}{\rho_{pp}(0)} = 
 \frac{2N_{\rm{ch}}}{N_{\rm{part}}\, N_{\rm{ch}}^{pp}} 
 \, \sqrt{\frac{L_{pp}}{L_{NN}}} \, .
 \label{eqn1}
 \end{equation}
  In Eq.(\ref{eqn1}) 
 the relation of the pseudorapidity 
 density and the mean multiplicity 
 is applied in its Gaussian form as obtained in Landau hydrodynamics.
 The 
 factor $L$ 
 is 
 defined 
 as 
  $L =  
 {\ln}({\sqrt{s}}/{2m})$.
  According to the 
 approach considered, 
 $m$ 
 is 
 the proton mass, $m_{p}$, in nucleus-nucleus 
collisions 
 and 
 the constituent quark mass in $pp/\pbp$ collisions which is set 
to 
 $\frac{1}{3}m_{{p}}$.
 $N_{\rm{ch}}$ and 
 $N_{\rm{ch}}^{pp}$ 
are the mean multiplicities in nucleus-nucleus and nucleon-nucleon 
collisions, respectively,  
and
 $N_{\rm {part}}$ is the number of participants. 
 Then,
 one 
 evolves
 Eq.~(\ref{eqn1}) 
 for the rapidity 
 density $\rho(0)$ and the multiplicity $N_{\rm{ch}}$ at $\sqrt{s_{NN}}$, 
and the rapidity 
density $\rho_{pp}(0)$ and the multiplicity $N_{\rm{ch}}^{pp}$ at $3 
\sqrt{s_{NN}}$: 

\begin{eqnarray}
\nonumber
&\rho(0)& = \rho_{pp}(0)\, 
\frac{2N_{\rm{ch}}}{N_{\rm{part}}\, N_{\rm{ch}}^{pp}} 
\, \sqrt{1 - 
\frac{4 \ln 3}{\ln (4m_p^{2}/s_{NN})}}\,,  \\
    & & \sNN=\spp/3 \, .
\label{eqn3}
\end{eqnarray}


 It was found \ct{edward} that 
 the current approach is able to reproduce very well 
the 
 data on  the 
 c.m. energy 
dependence of the midrapidity density 
measured in 
the most central heavy-ion collisions 
 by
 interrelating by  
Eq.~(\ref{eqn3}) 
the measurements in hadronic 
and 
nuclear collisions up to the top RHIC energy. Moreover, 
it was also shown that 
similarly, 
the total 
multiplicities in these types of collisions follow the  
 energy-dependence universality. 
 Furthermore, the proposed factor 1/3
 allows to relate {\it both} the 
 multiplicity and midrapidity c.m. energy dependence in $e^+e^-$ 
 and $pp/\pbp$ interactions and  solves 
 the problem of the factor 1/2, the latter been  introduced in 
\ct{phobos-rev}
 to account for
 the half 
 of the 
energy
 lost 
 attributed to
the leading protons.
 If the factor 1/2 is found to lead  
 to
 some
 similarity in the
 multiplicity data, 
 it cannot encompass
 the comparison of the midrapidity density.
 Interestingly, 
  the 
 3NLO 
 perturbative  QCD \ct{3nlo} fit
 to 
 $e^+e^-$
 data has been shown  \ct{pprev} to describe the multiplicity measurements 
in
  $pp/\pbp$  interactions up to TeV energies
  provided the inelasticity is set to $\approx 0.35$, \ie\ the effective
  1/3 energy in hadronic interactions. 
 Earlier, the  
  factor
  1/3 has been already shown to provide an agreement in 
  $e^+e^-$ 
   and
  $pp/\pbp$
  mean multiplicity data \ct{lep}. 
 Such a universality is found to 
 correctly predict
 \ct{edward} 
 the value of the 
 midrapidity density
  in {\it pp} interactions at
  the
  TeV LHC energies
  \ct{cms-conf}. 

In this paper, we 
extend the above-discussed 
 approach
 of the 
constituent quark participants and Landau hydrodynamics to 
the midrapidity pseudorapidity density dependence on the number of 
(nucleon) 
participants. Based on this energy dissipation picture, we apply effective 
energy consideration to 
the pseudorapidity density of the transverse energy 
at midrapidity, namely 
to the dependence of this observable on the c.m. energy and on the  
number of
participants measured in heavy-ion collisions in the RHIC and  LHC 
experiments. We give 
predictions for foreseen higher-energy heavy-ion collisions at the LHC.
\\


\begin{figure*}\sidecaption
\resizebox{0.6\textwidth}{!}{%
  \includegraphics{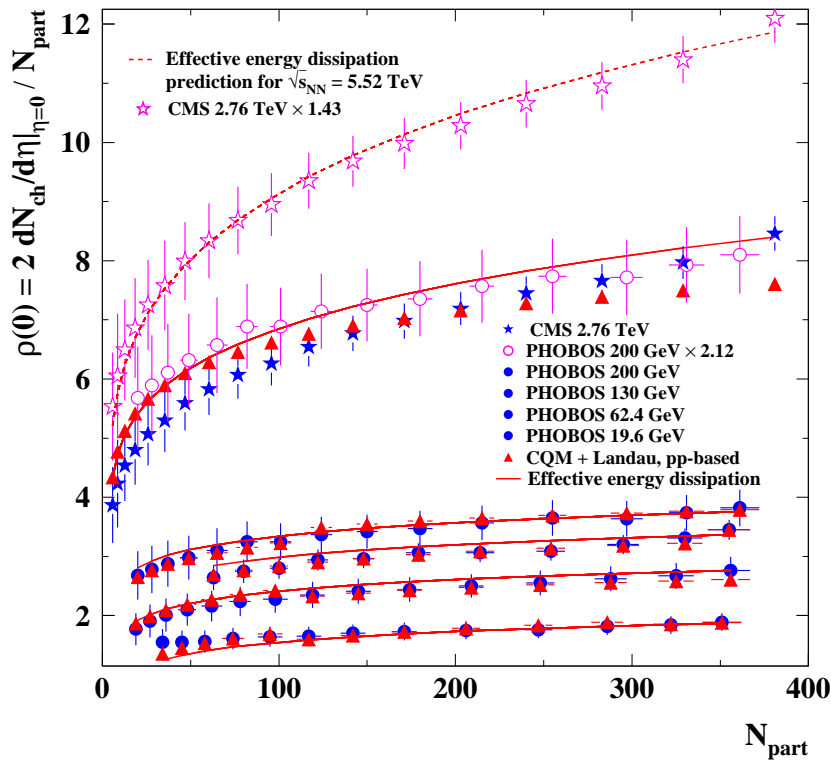}}
\caption{
 The 
 charged particle pseudorapidity density 
at midrapidity 
per participant 
pair as a function of the number of participants, $N_{\rm part}$.
 The 
 solid 
 circles show the dependence 
measured in
AuAu collisions at RHIC by 
PHOBOS at $\sNN=19.6$ to 200~GeV  
\ct{phobos-all} (bottom to top). The 
 solid
 stars show the measurements 
in 
PbPb collisions at LHC by CMS at $\sNN=2.76$~TeV  \ct{cms276c}.
 The 
 solid 
 triangles show the calculations by 
Eq.~(\ref{eqn4}) using $pp/\pbp$  data. 
 The 
 lines represent the effective energy dissipation 
approach predictions    
based on the hybrid fit
 to the c.m. energy
dependence of the 
midrapidity density in central heavy-ion collisions shown in 
Fig.~\ref{Fig5}.
 The 
 open 
 circles show the PHOBOS measurements at $\sNN=200$~GeV 
multiplied by 
2.12, while the open 
 stars show the CMS measurements multiplied by 
1.43.
 }
\label{Fig2}       
\end{figure*}

\begin{figure*}
\resizebox{0.63\textwidth}{!}{%
\includegraphics{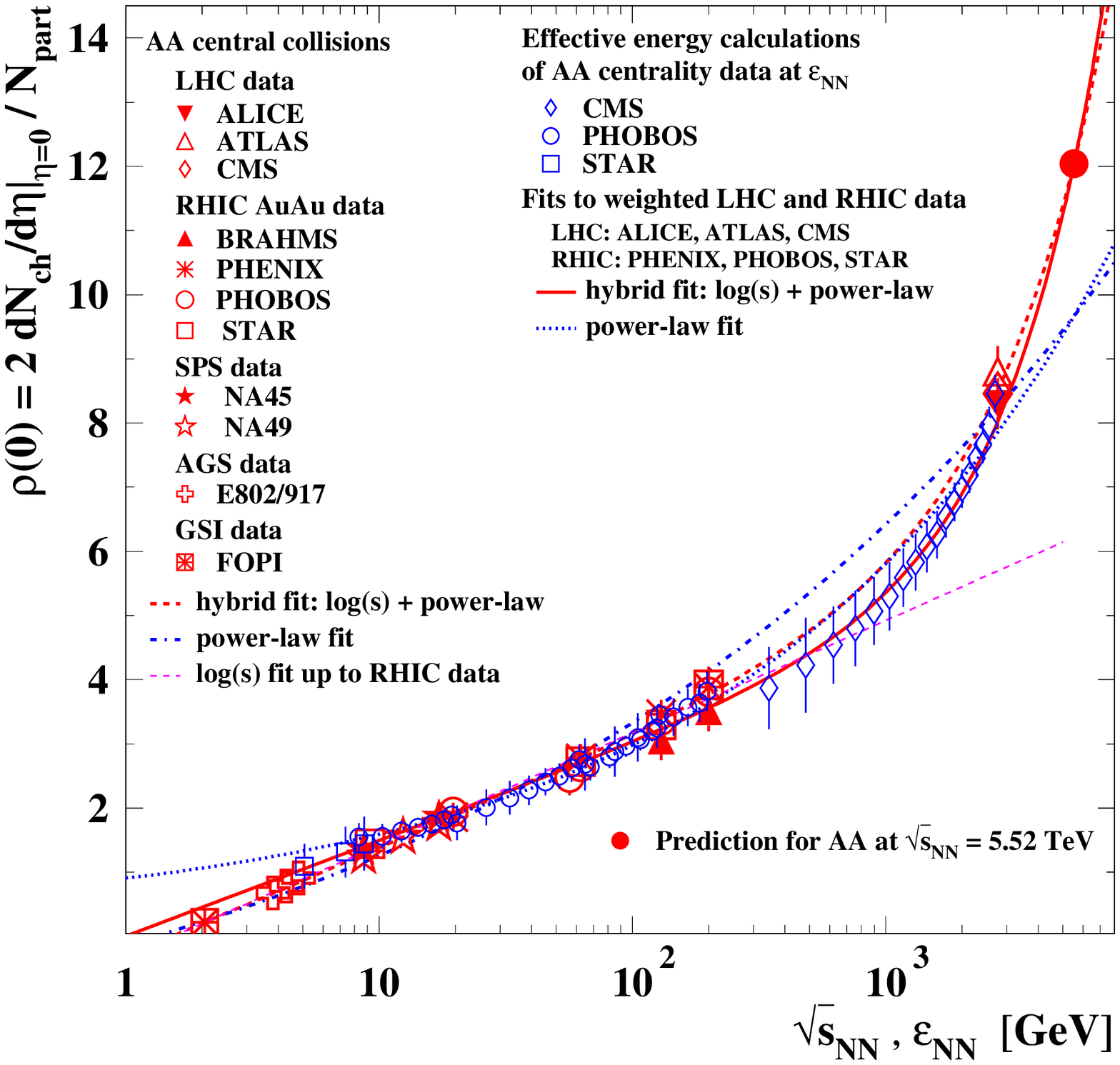}}
 \caption{The 
charged particle
pseudorapidity density  
 per participant
pair at midrapidity as a function of c.m. energy per nucleon,  $\sNN$, in 
central nucleus-nucleus (AA) collisions (shown by large 
 symbols), and  
as a 
function of effective c.m. 
energy, $\eNN$ (Eq.~(\ref{Eeff})), for AA collisions at 
 different 
 centrality (small 
 symbols).
 The data of central AA collisions  are from: the PbPb measurements at 
LHC by ALICE \ct{alice276r}, ATLAS \ct{atlas276c}, and CMS \ct{cms276c} 
experiments; 
the AuAu measurements
at RHIC by BRAHMS \ct{brahms}, PHENIX \ct{phenix-all}, PHOBOS 
\ct{phobos-all}, 
and STAR \ct{star-all,star9.2} experiments; the values recalculated
in \ct{phenix-all} from the measurements at CERN SPS by CERES/NA45 
\ct{na45}
and NA49 \ct{na49} experiments, at Fermilab AGS by E802 and E917 
experiments
\ct{ags}, and at GSI by FOPI Collab. \ct{fopi}.
 The centrality data represent the measurements by 
CMS \ct{cms276c}, 
PHOBOS \ct{phobos-all}, 
and 
STAR \ct{star9.2}; the CMS and PHOBOS data 
are those from Fig.~\ref{Fig2}, while for clarity, just every second point 
of the PHOBOS measurements  is shown. 
The 
 dashed-dotted 
 line and the dashed 
 line show the fits to the 
central collision data: 
the power-law fit, $\rho(0)=-2.955 +2.823\, \sNNq^{0.087}$, and 
the hybrid fit, $\rho(0)=-0.306+ 0.364\ln(\sNNq) +0.0011\,  \sNNq^{0.5}$.
The thin dashed 
 line shows the linear log fit, $\rho(0)=-0.327 
+0.381\,   
\ln(\sNNq)$ \ct{edward} to 
the central collision data up to the top RHIC energy. 
The   
 dotted 
 line and the solid 
line show the fits to the centrality data: the power-law fit,  
 $\rho(0)=0.244 +0.663\, \eNN^{0.308}$, 
and the 
hybrid fit, 
 $\rho(0)=0.002+ 0.646 \ln(\eNN) +0.0003 \, \eNN^{1.158}$,
respectively.
The fitted
centrality data
 include, 
except of the shown data,  also the 
  measurements  by  
ALICE \ct{alice276c} and ATLAS \ct{atlas276c}  at the LHC, and by PHENIX 
\ct{phenix-all}
and STAR \ct{star-all,star9.2} at RHIC (not shown).
 The solid 
 circle shows the prediction for $\sNN=5.52$~TeV.
}
\label{Fig5}       
\end{figure*}



{\bf 2. } 
  In Fig.~\ref{Fig2}, 
 the   charged particle 
pseudorapidity 
density per 
 participant pair at midrapidity as a function of number of participants 
 is shown as   measured by PHOBOS experiment
in AuAu 
 collisions at RHIC at 
 c.m. energy of $\sNN=19.6$ to 200~GeV \ct{phobos-all} 
 and by CMS experiment in PbPb collisions at LHC 
 at $\sNN=2.76$~TeV \ct{cms276c}, 
respectively. The PHOBOS data at  $\sNN=200$~GeV 
multiplied by 2.12 
are also shown to allow comparison with the LHC data and the 
 current
calculations. 
 As it is noted above, this dependence cannot be reproduced by the 
wounded nucleon model where a  number-of-nucleon-participant scaling 
is expected. 

 Within the above-discussed model of constituent quarks and Landau 
hydrodynamics, we consider this dependence in terms of centrality. 
 The centrality is considered to characterize 
 the degree 
 of overlapping 
 of the 
 volumes
  of the 
two colliding 
nuclei, determined by the impact parameter. The most central collisions 
correspond therefore to the lowest centrality while the larger
centrality define more peripheral collisions. The centrality is closely 
related to the number of nucleon participants determined using a Monte 
Carlo Glauber calculations so that the largest number of participants 
contribute to the most central heavy-ion collisions. Hence the centrality 
is related to the energy released in the collisions, \ie\ 
the effective energy, $\eNN$, which, in the framework of the 
 proposed approach,
 can be 
defined as 
a fraction of the c.m. energy available in a collision 
according to the centrality, $\alpha$: 

\begin{equation}
\eNN = \sqrt{s_{NN}}(1 - \alpha).
\label{Eeff}
\end{equation}
 Conventionally, the data are divided into classes of centrality, or 
centrality intervals, so that $\alpha$ is the average centrality for the 
centrality interval, 
\eg\  $\alpha = 0.025$ for 
$0-5\%$ centrality, which refers to the 5\% most central collisions.
 In what follows we have  checked
 that for a particular centrality interval the 
 conclusions and the 
 results are not influenced
 by taking 
 either the 
 mid-point of the centrality interval or both the extremes.


  In fact, each of the scalings described by Eq.(\ref{eqn3}) and 
Eq.(\ref{Eeff}) regulates a
particular physics ingredient used in the modeling of our approach.
Namely, the scaling introduced by Eq.(\ref{eqn3}) embeds the constituent 
quark 
model
which leads to establishing a similarity between hadronic and 
nuclear collisions, 
while
the scaling driven by Eq.(\ref{Eeff}) is appealed to define the energy 
budget
effectively retained for multiparticle production in the most central 
collisions
to determine the variables obtained from centrality data.

Then, for 
 the 
effective
c.m.  energy $\eNN$, 
   Eq.~(\ref{eqn3})  reads:
 
\begin{eqnarray}
 \nonumber
 &\rho(0)&= 
 \rho_{pp}(0)\, \frac{2N_{\rm ch}}{N_{\rm{part}}\, N_{\rm{ch}}^{pp}} 
\, \sqrt{1 - 
\frac{2 \ln 3}{\ln (2m_p/{\eNN})}}\, ,
\\
& & \eNN=\spp/3\,,
\label{eqn4}
\end{eqnarray} 
where
$N_{\rm{ch}}$ 
is the  mean multiplicity in central nucleus-nucleus collisions measured 
at  $\sNN=\eNN$.
 The 
 rapidity 
 density $\rho_{pp}(0)$ and the multiplicity $N_{\rm{ch}}^{pp}$ 
 are taken from the existing data or, where not available, calculated 
using the  
 corresponding experimental c.m. energy fits\footnote {The E735 
power-law fit
$N_{\rm{ch}}^{pp}
= 3.102\,s_{pp}^{0.178}$ \ct{pprev} is used, while 
the 
linear log fit 
$\rho_{pp}=-0.308+0.276\, \ln(s_{pp})$ \ct{pprev} and the 
 power-law fit by CMS \ct{cms276c},  $\rho_{pp}=-0.402+s_{pp}^{0.101}$,
 are
 used for $\sqrt{s_{pp}}\leq$~53~GeV and for $\sqrt{s_{pp}}>$~53~GeV, 
respectively.}, 
  and, according to the 
 consideration,
  the calculations are made at $\spp=3\, \eNN$. 
The $N_{\rm ch}$ values are as well 
taken from the 
measurements in central heavy-ion collisions wherever available, while for 
the non-existing data the 
``hybrid" 
fit \ct{aditya} combining the linear logarithmic and power-law 
regularities  is used.  
 This fit is inspired by the measurements  as well as by theoretical 
considerations. It is observed that the logarithmic fit 
well 
describes the heavy-ion multiplicity data up to the top RHIC energy
\ct{edward,atlas276c}, 
however as 
 the collision energy 
increases above 1$-$2 TeV at the LHC, the data clearly show a preference 
for the power-law 
behaviour \ct{cms276c,atlas276c,alice276r} in the multiplicity dependence 
 on
$\sNN$.
   From the theoretical description point of view, such a c.m. energy 
dependence is 
expected~\ct{wolschin} as soon as
the logarithmic dependence is considered to characterize the fragmentation 
source(s) 
while the power-law behaviour is believed to come from the gluon-gluon 
 interactions.
 

 In the framework of the model of constituent quarks 
 combined with
 Landau hydrodynamics, 
we calculate the centrality dependence of the charged particle 
midrapidity density using Eq.~(\ref{eqn4}) to reproduce  
the 
centrality data shown in 
Fig.~\ref{Fig2}. The calculations are shown by solid triangles. 
 One can see that within this 
 approach
 where the collisions are  
derived by the centrality-defined effective c.m. energy $\eNN$,   
 the 
 calculations 
are in 
  very good overall agreement with the 
 measurements independent of the collision energy. Similar results are 
obtained as the $N_{\rm part}$-dependence of the PHENIX \ct{phenix-all}, 
STAR \ct{star-all}, or CuCu PHOBOS \ct{phobos-all} measurements  from 
RHIC and ALICE \ct{alice276c} or ATLAS \ct{atlas276c} data from LHC are 
used (not shown).
 Some slightly lower values are however seen in the 
 calculations
compared to the data  
for some low-$N_{\rm part}$, \ie\ for the most peripheral collisions,
at $\sNN=19.6$~GeV,  and for a couple of central data points obtained 
at the 
highest $\sNN$.
  The deviation observed in the peripheral collisions at $\sNN=19.6$~GeV
 looks to be 
due to the experimental limitations and the 
extrapolation used in the reconstruction  for
the measurements in this region of very low multiplicity \ct{phobos-all}. 
 This also  may explain the $N_{\rm part}$-scaling of the data at 
$\sNN=19.6$~GeV in the most peripheral region so the data of these 
 centrality intervals do not follow the common trend of decreasing  as it 
 is observed 
 in higher-energy measurements. 
 The low values obtained 
 within the approach
 for a few  most central collisions 
 at the LHC energy  can be explained 
 by no data 
 on  $N_{\rm ch}^{pp}$ 
 being 
 available 
at $\spp>1.8$~TeV.  Moreover, 
 for $\spp>53$~GeV,
the second-order 
logarithmic polynomial fit to the $\spp$ dependence of 
$N_{\rm ch}^{pp}$ 
is indistinguishable from the exponential function fit 
\ct{pprev}. The 
latter 
regularity is used here 
 for the $\spp$ dependence to calculate $N_{\rm ch}^{pp}$ 
above 
the Tevatron energy.  

Given the obtained agreement between data and the 
 calculations
 and 
considering the similarity put forward for  $\eNN$ and 
$\sNN$, one would expect the measured 
centrality data at $\eNN$ to follow the  $\sNN$ dependence of the 
midrapidity density in the most central nuclear collisions. In 
Fig.~\ref{Fig5}, the measurements of the charged particle 
pseudorapidity 
density at midrapidity 
 in head-on nuclear collisions are plotted against 
the 
$\sNN$ from a few GeV at GSI to a few TeV at the LHC
 along with the 
centrality data, shown as  a
function of $\eNN$,  from low-energy RHIC data by STAR at 9.2 GeV 
\ct{star9.2}, and the measurements, shown in Fig.~\ref{Fig2}, by PHOBOS 
\ct{phobos-all} and 
CMS \ct{cms276c} experiments
 as a 
function of $\eNN$. 
 The centrality data effective-energy dependence follow well the 
data on the most central collision c.m. 
energy behaviour.
 
 We fit the  weighted combination of the midrapidity density from 
the head-on collisions by the hybrid fit function 
\begin{eqnarray}
\nonumber
 \rho(0)& = & (-0.306\pm 0.027)+ (0.364\pm 0.009)\,
 \ln(\sNNq)\\ 
& & +(0.0011\pm 
0.0011) \,  \sNNq^{(0.50\pm 0.06)},
\label{hybrho} 
\end{eqnarray}
which is, as it is noticed above, inspired by the measurements and 
supported by theoretical consideration. The fit combines the linear-log 
dependence 
on $\sNN$  observed up to the top RHIC energy \ct{phobos-all,phenix-all} 
and the power-law dependence obtained with the LHC data \ct{cms276c,atlas276c,alice276r}. This fit is shown in 
Fig.~\ref{Fig5} by the dashed 
 line. One can see that the fit is as 
well close to the 
centrality data.  To clarify, the weighted combination of the centrality 
data are also 
fitted 
by the hybrid function, 
\begin{eqnarray}
\nonumber
  \rho(0)& = & (0.002\pm 0.080)+ 
 (0.646\pm 0.022)\, \ln(\eNN)\\ 
 & & +(0.0003\pm 0.0001) \,  
 \eNN^{(1.158\pm 0.034)},
\label{hybrhoc} 
\end{eqnarray}
 where, in addition to the low-energy STAR 
data and the measurements, shown in Fig.~\ref{Fig2}, by 
the PHENIX and CMS experiments, 
 the midrapidity density data on the  centrality dependence from 
ALICE \ct{alice276c}, ATLAS \ct{atlas276c}, PHOBOS \ct{phobos-all} and 
STAR 
\ct{star-all} are included (not shown). 
 The fit is shown by the solid 
 line and is very close to the fit made 
to the 
 head-on collision
 data. From this one can conclude that the 
 picture proposed
 well 
reproduces the data under the assumption of the effective energy 
deriving the multiparticle production process pointing to the 
similarity in all the data from peripheral to the most central 
measurements
to 
follow 
the same energy behaviour.  
 From the fit, we estimate the midrapidity density value to be of about 
12.0  within 10\% uncertainty in the 
most central collisions at $\sNN=5.52$~TeV shown by the solid 
 circle 
in Fig.~\ref{Fig5}. 

 In addition to the hybrid fits, in Fig.~\ref{Fig5} we show the 
linear-log fit  \ct{edward} 
up to the top 
RHIC energy (thin 
 dashed 
 line) 
and the 
power law fit for the entire energy range  
 (dashed-dotted 
 line) 
to  the most 
central  
collision data along with  
the power law fit to the centrality data 
 (dotted 
 line). All the 
fits are made 
 by using 
the weighted data as above. One can see that the power-law fit 
describes well the head-on collision measurements (see also \ct{advrev}) 
and, within the errors, 
does 
not differ from the 
linear-log or the hybrid functions up to the RHIC energies. However it 
 deviates from 
the  most central collision  hybrid fit 
  as soon as the LHC measurements are included. The power-law fit to 
the 
centrality data are much closer to the hybrid fits, and it is almost 
indistinguishable from the hybrid fit to the central data up to the 
head-on collision LHC points. 
Both the power-law fits, to the head-on 
collision  data and to the 
 centrality data,  give predictions close to each other but lower than the 
hybrid fits   
up to some higher c.m. energies.    
 Interestingly, using the approach of the effective energy dissipation, 
one can clearly see the transition to a possibly new regime in the 
multihadron production in heavy-ion collisions demonstrated by the data as 
$\sNN$ 
increases up to 
about 600--700 GeV per nucleon. The 
centrality data still follow the central collision data and the log 
fit up to these energies while then the energy behaviour changes to the 
 power-law  one. 
 The change in the $\sNN$-dependence from the logarithmic to the 
power-law one seems to be a reason of lower-value predictions by theoretical 
models 
 \ct{alice276r}. The change also restrains predictions for heavy-ions 
 within the universality  
 picture \ct{edward} which however gives the correct predictions for 
$pp/\pbp$ \ct{cms-conf}, where 
 both the logarithmic \ct{cmspp7pTn} and the power-law \ct{cms276c} functions provide equally good fits to the data 
up to $\spp=7$~TeV.

Now, using the effective c.m. energy approach, we apply the obtained hybrid function fit of the 
midrapidity density measured in {\it head-on}  collision data, Eq.~(\ref{hybrho}), to the {\it centrality} data,
shown in Fig.~\ref{Fig2} as a function of $N_{\rm part}$. The calculations are shown by the solid 
 lines. One 
can see that the 
 approach
 well describes the measurements and actually follows the 
predictions by Eq.~(\ref{eqn4}), 
except the LHC data, where it is better than the calculations of 
Eq.~(\ref{eqn4}), though slightly  overshoots 
 the measurements. Similar to the 
 consideration
 combining constituent quarks and 
Landau hydrodynamics, the calculations using the effective energy $\eNN$ 
show lower values for the very peripheral points at the lowest c.m. 
energy, $\sNN=19.6$~GeV.  The 
difference, as mentioned above, seems to be due to the 
difficulties in the 
measurements because of the very low multiplicity in these data. A slight 
overestimation of the LHC data is due to the fact that the fit 
(Eq.~(\ref{hybrho})) uses the highest (0-2\% centrality) ATLAS  point of 
the 
head-on collisions.

Similarly to the above calculations for the existing data on the $N_{\rm 
part}$-dependence of the midrapidity density, we made the 
predictions for the forthcoming heavy-ion collisions at $\sNN=5.52$~TeV. 
The predictions are shown by the dashed 
 line in Fig.~\ref{Fig2}, where 
the centrality and $N_{\rm part}$ values are taken as in the 2.76~TeV 
data shown. The expectations show increase of the $\rho(0)$ 
with $N_{\rm part}$ (decrease with centrality) from about 5 to 12. The 
increase  looks to be 
faster than at $\sNN=2.76$~TeV, especially for the peripheral region, 
similar to the change in the behaviour seen as one moves from the RHIC 
measurements to the LHC data, cf. 200 GeV data and 2.76 TeV data in 
Fig.~\ref{Fig2}. We find that the predictions made here are well 
reproduced when 
the LHC data are scaled by a factor 1.43, similar to the
multiplication factor (of 2.12 shown here) found 
\ct{cms276c,atlas276c,alice276c} to reproduce the 2.76 TeV LHC data by 
the 200 GeV RHIC ones.    

 Interestingly, within the picture of the effective energy dissipation of 
constituent quark participants one can explain the observed 
similarity of the midrapidity densities measured in $pp/\pbp$ interactions 
and 
in heavy-ion collisions at the same c.m. energy, as soon as in the latter 
case the data are 
recalculated in  the 
constituent quark framework \ct{indpart,nouicer}. Moreover, this approach 
supports the scaling with the number of partcipants of the midrapidity 
pseudorapidity and transverse energy densities 
obtained for RHIC \ct{eremin,ind,indpart,nouicer,phenixEt} and LHC 
\ct{aditya} data in the constituent quark 
framework. Note that  this 
scaling been  observed also for most peripheral 
collisions may be understood in the framework of the approach proposed 
here by considering the most peripheral collisions to be driven by 
nucleon-nucleon interactions where a pair of participating constituent 
quarks each per nucleon contribute, thus the fraction of c.m. energy, \ie\ 
the  
effective energy  of the 
participants is pumped into the small collision zone of the overlapped 
nuclei.    
\\


 


\begin{figure*}
\resizebox{0.63\textwidth}{!}{%
\includegraphics{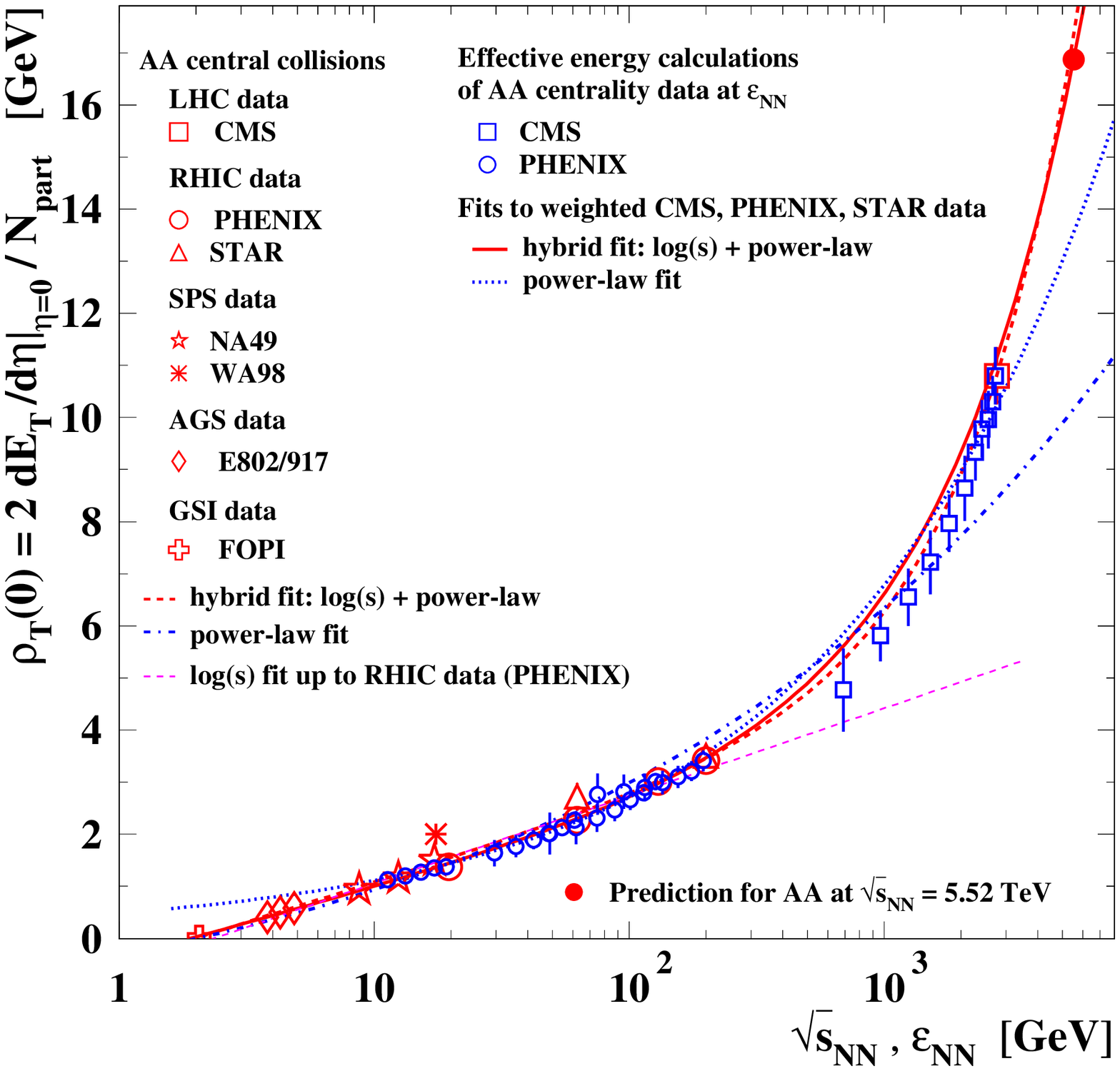}}
\caption{The 
charged particle transverse energy
pseudorapidity density 
 per participant
pair at midrapidity as a function of c.m. energy per nucleon,  $\sNN$, in 
central nucleus-nucleus (AA) collisions (shown by large 
 symbols), and  
as a 
function of effective 
energy, $\eNN$ (Eq.~(\ref{Eeff})), for AA collisions at different 
centrality (small 
 symbols).
 The data of central AA collisions  are from: the PbPb measurements at 
LHC by CMS \ct{cms276Et} 
experiment; 
the AuAu measurements
at RHIC by PHENIX \ct{phenix-all,phenixEt} 
and STAR \ct{star200EtRS62.4} experiments; 
the values recalculated
in \ct{phenix-all} from the measurements
at CERN SPS by CERES/NA45 
\ct{na49}
and WA98 \ct{wa98Et} experiments, at Fermilab AGS by E802 and E917 
experiments
\ct{agsEt}, and at GSI by FOPI Collab. \ct{fopi}.
 The centrality data represent the measurements by 
CMS at the LHC \ct{cms276Et} 
and by PHENIX at 
RHIC
 \ct{phenix-all,phenixEt}; the CMS and PHENIX data 
are those from Fig.~\ref{Fig6}, while for clarity, just every second point 
of the PHENIX measurements  is shown. 
The dashed-dotted 
 line and the dashed 
 line show the fits to the 
central collision data: 
the power-law fit, $\rho_T(0)= -2.29 +1.97 \sNNq^{0.107}$, 
 and 
the hybrid fit, 
 $\rho_T(0)=-0.447+ 0.327\ln(\sNNq) +0.002\,  \sNNq^{0.5}$.
The thin dashed 
 line shows the linear log PHENIX fit 
 \ct{phenix-all} 
to 
the central collision data up to the top RHIC energy. 
The 
dotted 
 line 
and
 the solid
 line 
show 
the fits to the centrality data:
the 
power-law fit,
 $\rho_T(0)=0.09+ 0.40\,\eNN^{0.40}$,
 and
 the hybrid fit, 
 $\rho_T(0)=-0.387+ 0.574\ln(\eNN)+ 0.011\, \eNN^{0.818}$, respectively.
The fitted
centrality data
 include, 
 except of the shown data,  also the 
  measurements  by   
 STAR \ct{star200EtRS62.4} at RHIC (not shown).
 The solid 
  circle shows the prediction for $\sNN=5.52$~TeV.
 }
\label{Fig7}       
\end{figure*}

\begin{figure*}\sidecaption
\resizebox{0.6\textwidth}{!}{%
\includegraphics{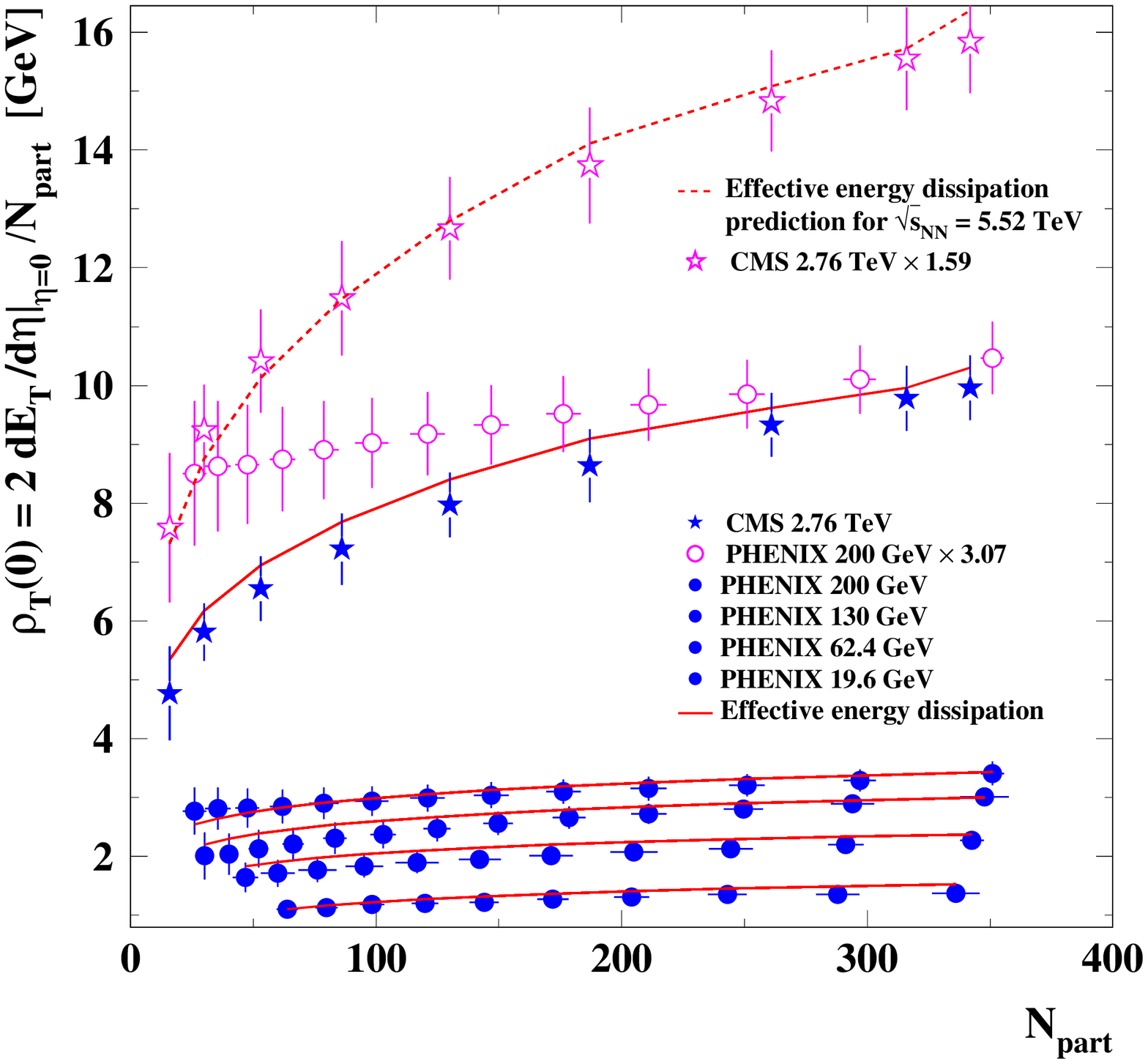}}
\caption{The charged particle 
 transverse energy pseudorapidity 
 density  
at midrapidity 
per participant 
pair as a function of the number of participants, $N_{\rm part}$.
 The solid 
  symbols 
show the data from 
AuAu collisions at RHIC (circles) by 
PHENIX experiment at $\sNN=19.6$ \ct{phenix-all} and 62.4, 130 and 
200~GeV 
\ct{phenixEt}  
 (bottom to top) 
 and  from
 PbPb collisions at LHC by CMS at $\sNN=2.76$~TeV  \ct{cms276Et} (stars). 
 The 
 lines
show the predictions by the effective energy approach 
 using 
 the hybrid fit
 to the c.m. energy
dependence of the 
midrapidity transverse energy density in central heavy-ion collisions  
shown in
Fig.~\ref{Fig7}.
 The open 
 circles show the PHENIX measurements at $\sNN=200$~GeV 
multiplied by 
3.07, while the open 
 stars show the CMS data multiplied by the factor 
1.59.
 }
\label{Fig6}       
\end{figure*}

{\bf 3. }  The effective c.m. energy approach applied to the 
charged particle pseudorapidity 
 density at midrapidity can be considered to be applied to another 
important variable, 
such as 
the pseudorapidity density  of the transverse energy, 
$\rho_T(\eta)=(2 /N_{\rm part}) \,dE_T/d\eta$, at 
midrapidity, 
 $\eta 
\approx$~0. The charged particle density and the transverse energy density 
are 
closely related  and, been  studied together, provide important
characteristics of the underlying dynamics of the multihadron production.  
 The transverse energy measurements, as well as the pseudorapidity data, 
have been shown to be reasonably well modelled by the constituent quark 
picture 
\ct{ind,aditya,phenixEt}. 

In Fig.~\ref{Fig7}, the $\sNN$ dependence of the  charged particle 
midrapidity transverse 
energy 
density in 
pseudorapidity  is displayed as measured in head-on collisions at 
the experiments from a few GeV at GSI to a few TeV at the LHC, shown by 
the 
 big 
 symbols. On top of 
these data the centrality data from the  PHENIX experiment at RHIC 
\ct{phenixEt} and the CMS experiment at LHC \ct{cms276Et} are added as 
a function of the effective c.m. energy $\eNN$  shown 
by small 
 symbols. Similarly to the case of the charged particle 
density at midrapidity, the $E_T$ density data show the complementarity 
of these two types of measurements: the centrality data follow  well the 
data from 
the central collisions. 

To better trace the similarity in the energy dependence of the central 
collision and the centrality-dependent data, we fit the data by the 
hybrid function, as is done in Fig.~\ref{Fig5} for the particle 
psudorapidity densities.
 For the central collisions one gets:
 \begin{eqnarray}
\nonumber
 \rho_T(0)&=&(-0.447\pm0.014)+ (0.327\pm 0.011)\, \ln(\sNNq)\\ 
& &+(0.002\pm0.003)\,  \sNNq^{0.50\pm0.08},
\label{hybet}
\end{eqnarray}
and similar fit to  the centrality data reads:
  \begin{eqnarray}
\nonumber
 %
 \rho_T(0)&=&(-0.387\pm0.090)+ (0.574\pm 0.032)\, \ln(\eNN) \\
 & &+(0.011\pm0.005)\,  \eNN^{0.818\pm 0.064}\,.
 \label{hybetc}
\end{eqnarray}
 The fits are shown in Fig.~\ref{Fig7} by the dashed 
  and solid 
 lines, 
respectively. The data from different experiments are weighted, and the  
fit of the effective c.m. energy $\eNN$ includes the STAR measurements  
in addition to the PHENIX ones. One can see that the two fits are 
amazingly 
close to each other for the entire energy range  
allowing to  conclude that the effective energy approach provides 
a good description of the $E_T$ production in heavy-ion collisions.
 We estimate the value of $\rho_T(0)$ to be about 16.9~GeV  with about 
10\% 
uncertainty for most central 
collisions at $\sNN=5.52$~TeV shown by the solid 
 circle in 
Fig.~\ref{Fig7}.  

  As it is obtained above  
for the midrapidity pseudorapidity density 
energy dependence (Fig.~\ref{Fig5}), in Fig.~\ref{Fig7} 
the LHC data demonstrate a clear departure from the linear-log
regularity in the region of $\sNN 
 \simeq
500-700$~GeV; the log fit to the data up to the top RHIC energy is 
shown by the thin  dashed 
 line and is taken from \ct{phenix-all}.
 This observation supports a possible transition to  a new regime in 
heavy-ion 
collisions at $\sNN$ above a few hundred GeV as indicated by the 
midrapidity density in Fig.~\ref{Fig5}. 
 In Fig.~\ref{Fig7}, we also show the power-law fits  to the central 
collision measurements by the 
 dashed-dotted 
 line and to the 
centrality data by the 
 dotted 
 line. As above, in the fits the data from 
different experiments are weighted. 
 One can see that the power-law fit to the central collision data 
underestimates the LHC measurement at 2.76 TeV data and deviates  from the 
data at  
$\sNN\sim 1$~TeV. However, the power-law fit to the centrality
describes well the data in the full available c.m.-energy 
region, though lies slightly lower than the 
hybrid   fit, Eq.~(\ref{hybetc}). Meantime, this fit overestimates the 
data 
below $\sNN \approx 10$~GeV, similarly to the case of the multiplicity 
data 
on 
centrality, Fig.~\ref{Fig5}. Interestingly, the shown  power-law fit curve 
to the 
centrality 
data is similar to that obtained by CMS for $\sNN\ge 8.7$~GeV 
\ct{cms276Et}; moreover, fitting all the $E_T$ {\it centrality} data  
  \`a-la 
CMS, one 
finds a good fit to the data by $\rho_T(0)=0.43 \,\eNN^{0.20}$ (not 
shown) which resembles 
the 
CMS fit, $\rho_T(0)=0.46 \,\sNNq^{0.20}$, to the {\it head-on} collision 
data. 
This 
again demonstrates  
the 
multihadron production in heavy-ion collisions to be well described by the 
effective c.m. energy dissipation 
 picture.

To further exploit the effective energy approach with the centrality data, 
in Fig.~\ref{Fig6} we show the $N_{\rm part}$ dependence of the centrality 
data
from Fig.~\ref{Fig7} along with the central collision data fit, 
Eq.~(\ref{hybet}), but as a function of the centrality-dependent 
c.m. 
effective energy $\eNN$.
 One can see that the fit well describes the data; in this case the 
agreement is even better than 
for the midrapidity density,
as one concludes from the comparison with the LHC centrality data.  
Interestingly, 
the 
 open 
 circles which represent the RHIC data at $\sNN=200$~GeV scaled by 3.07, 
to allow 
comparison with the LHC measurements, demonstrate much less decrease as 
the 
centrality increases (more peripheral data), than that 
observed for the LHC data. This is different for the pseudorapidity 
density of charged particles at midrapidity measurements, 
see Fig.~\ref{Fig2}. In contrast to 
the scaled RHIC data, the effective energy approach follows well
 the LHC measurements. 
 %

 Similarly to the above comparison to the existing data on the $N_{\rm 
part}$-dependence of the midrapidity transverse energy density, we make 
the 
predictions for the future heavy-ion collisions at $\sNN=5.52$~TeV within 
the effective energy dissipation approach. 
The predictions are shown by the dashed 
 line in Fig.~\ref{Fig6}.
The predictions show more rapid increase 
 of the $\rho_T(0)$ 
with 
$N_{\rm 
part}$ (decrease with
centrality)  
 than at $\sNN=2.76$~TeV, especially for the peripheral region, 
similar  
to the change in the behaviour seen as one moves from the RHIC 
measurements to the LHC data and similar to that obtained for the 
 midrapidity density, 
Fig.~\ref{Fig2}. We find that the predictions made are well reproduced as 
the LHC data are scaled by a numerical factor 1.59, as shown  by 
open 
 stars, Fig.~\ref{Fig6}.


{\bf 4. }
   %
   In summary, we analyzed the midrapidity pseudorapidity density of 
charged particles 
   and of the transverse energy measured in nucleus-nucleus collisions in 
the whole 
available range of the collision c.m. energy per nucleon, $\sNN$, from a 
few GeV at 
GSI up to 
a few TeV at the LHC. The dependencies of these key variables on the c.m. 
energy per nucleon and on the number of participants (or centrality) 
 have been revealed
 within 
the approach 
of the dissipation of the effective energy pumped in by the participants 
of the 
 collisions,
 which forms the effective energy budget in the multiparticle production 
 process.
 Namely, the 
 model 
of constituent quarks combined with Landau 
hydrodynamics is applied to reproduce the midrapidity density 
dependence on the number of participants. This 
 approach,
 proposed earlier in \ct{edward} and pointed to 
the universality of the 
multihadron production in different types of collisions up to the top RHIC 
 energy 
 allows to well predict the LHC measurements in 
$pp/\pbp$ interactions on the midrapidity density of charged particles. 
Within this 
 picture,
 we find that the dependence of the pseudorpaidity 
density at midrapidity from the RHIC to LHC data is well reproduced as 
soon as the effective c.m. energy variable is introduced as the 
centrality-defined fraction of the collision c.m. energy. Based on this 
finding, it is shown that the most central collision data and the 
centrality-dependent data follow a similar $\sNN$ dependence obtained 
for the central collision data as soon as the centrality data is rescaled 
to the effective energy. The hybrid fit,
combining the linear-log and the power-law c.m. energy dependencies of
 the head-on collision data, where the 
linear-log function known to fit the measurements up to the top RHIC 
energy and the 
power-law regularity is needed up to the TeV LHC data, is found to well 
reproduce the 
dependence of the midrapidity densities on the number of participants 
within the effective energy approach. Similar observations are made for 
the 
transverse energy midrapidity density measurements: as soon as the 
centrality data is recalculated for the c.m. effective energy, these 
measurements are found to well complement the central collision data c.m. 
energy behaviour. The hybrid fit made to the central collision data is 
shown to reproduce well the midrapidity transverse energy dependence 
on 
the 
number of participants. 
 For both the variables studied, a 
clear departure of the data as a function of the effective c.m. energy
from the linear-log dependence to the power-law one
is observed  
at $\sNN 
 \simeq
 500-700$~GeV
 indicating a possible transition to a new regime 
 in heavy-ion collisions. The data at 
 $\sNN\sim 1$~TeV would be extremely useful to clarify the 
observations made 
here. 
Based on the hybrid fits in the framework of the discussed 
 approach,
 the predictions for the energy and the number of participant 
dependencies
 for the 
 measurements 
in the forthcoming heavy-ion runs at LHC 
 at
 $\sNN=5.52$ TeV are made.

 \vspace{0.5cm}
\noindent

%
%

\end{document}